\documentclass[12pt]{article}
\usepackage{scicite}
\usepackage{times}
\usepackage{graphicx}

\newenvironment{sciabstract}{%
\begin{quote} \bf}
{\end{quote}}

\newcounter{lastnote}

\title{Imaging resonances in low-energy NO-He inelastic collisions}

\author
{Sjoerd N. Vogels$^{\ast}$, Jolijn Onvlee$^{\ast}$, Simon Chefdeville, Ad van der Avoird, \\
Gerrit C. Groenenboom$^{\ast\ast}$, Sebastiaan Y.T. van de Meerakker$^{\ast\ast}$\\  \\
\\
\normalsize{Radboud University, Institute for Molecules and Materials}\\
\normalsize{Heijendaalseweg 135, 6525 AJ Nijmegen, the Netherlands}\\
\normalsize{$^{\ast}$Who contributed equally to this work;}\\
\normalsize{$^{\ast\ast}$To whom correspondence should be addressed;}\\
\normalsize{E-mail: basvdm@science.ru.nl, gerritg@theochem.ru.nl}}

\date{}

\begin{document}

\baselineskip24pt

\maketitle

\begin{sciabstract}
In molecular collisions, resonances occur at specific energies where the colliding
particles temporarily form quasi-bound complexes, resulting in
rapid variations in the energy dependence of scattering cross sections.
Experimentally, it has proven challenging to observe such scattering resonances,
especially in differential cross sections.
We report the observation of resonance fingerprints in the
state-to-state differential cross sections for inelastic NO-He
collisions in the 13 to 19 cm$^{-1}$ energy range with 0.3 cm$^{-1}$
resolution. The observed structures were in excellent agreement with
quantum scattering calculations. They were analyzed by separating the
resonance contributions to the differential cross sections from the
background through a partitioning of the multichannel scattering
matrix. This revealed the partial wave composition of the resonances,
and their evolution during the collision.
\end{sciabstract}%

\newpage

Acquiring a detailed understanding of interactions between individual
molecules is of fundamental importance to physics and chemistry, and has
a long and rich tradition. Historically, two distinct types of
experiments have been conducted to unravel the precise nature of inter-
and intramolecular forces. On the one hand, spectroscopic studies yield information on
the binding forces that hold the molecule together, as well as on the
noncovalent interactions between the molecules. They mainly probe the
shape of the interaction potential in the region of the well. Collision
experiments, on the other hand, are typically more sensitive to the
short range repulsive part of the potential.

Certainly, the interactions between microscopic objects such as atoms
and molecules require a full quantum mechanical description. In contrast
with molecular spectra, where the discrete lines and bands directly
reveal the quantized energy levels of the molecule, the quantum
nature of molecular collisions is more subtle. In addition to the
internal states of the molecule, the angular momentum associated with
the relative motion of the particles is also quantized. This relative
motion is described by a set of partial waves with integer quantum
number $\ell$ \cite{Note:4}.

Unraveling the contribution of each individual partial wave to a
collision cross section would provide the ultimate information that can
be retrieved from any collision event. Experimentally, however, it is
impossible to select a single partial wave from the
pre-collision conditions, and to study how the interaction transforms it
into post-collision properties. The number of contributing partial waves
depends on the de Broglie wavelengths of the particles; observable
quantities such as scattering cross sections therefore necessarily
represent the quantum mechanical superposition of many partial waves
\cite{Note:5}. Classically, this can be compared to the unavoidable
averaging over all possible impact parameters of the two colliding
particles \cite{Note:4}.

At low temperatures, special conditions exist where a single partial
wave can dominate the scattering process, mitigating this fundamental
obstacle. When the collision energy is resonant with a quasi-bound
state supported by the interaction potential, the incident
particles can temporarily form a long-lived complex. At these energies,
a resonant partial wave $\ell_{\rm res}$ may dominate over all other
partial waves, and there will be a strong enhancement of the scattering
cross section. For atom-molecule collisions, these so-called scattering
resonances may be regarded --- in a simplified picture --- either as the orbiting of the atom around the
molecule (a shape or orbiting resonance), or as the transient
excitation of the molecule to a state of higher energy (a Feshbach
resonance). After some time, however, the complex falls apart and decays
back into a separate atom and molecule \cite{Chandler:JCP132:110901}.

Scattering resonances are the most global and sensitive probes
of molecular interaction potentials. They depend on both the
long range attractive and the short range repulsive part of the
potential, as well as on the van der Waals well; they are not only
sensitive to the shape of the well --- as are the spectra of molecular
complexes --- but also to the depth of the well relative to the
dissociation limit. Measurements of the resonance position and width in
the integral cross section (ICS) probe the energy and lifetime
of the quasi-bound state from which the resonance originates. Such
observations may thus be regarded as a type of collision
spectroscopy.  Still, they do not yield information on the partial
wave composition of the resonance. More information on the collision
dynamics is inferred from the differential cross section (DCS) at the
resonance energy. The structured DCS represents the partial wave
fingerprint of the collision process, and offers at the resonance the
opportunity to experimentally probe the relation between incoming, resonant, and
outgoing partial waves. This gives detailed insights into the most
fundamental question in molecular collision dynamics: how does the
interaction potential transform the reagents into the collision
products?

Whereas scattering resonances are well-known in electron, neutron and
ultracold atom scattering, the observation of scattering resonances in
molecular systems has been a quest for decades. Experimentally, it has
proven extremely challenging to reach the required low collision
energies and high energy resolution to observe and characterize partial
wave resonances. In crossed beam experiments, signatures of scattering
resonances have been observed in reactive systems with a low excitation
barrier, such as the benchmark F + H$_2$ and F + HD reactions, using the
powerful Rydberg tagging technique to record the angular distribution of
the H or D products
\cite{Skodje:PRL85:1206,Shiu:PRL92:103201,Qiu:Science311:1440,Ren:PNAS105:12662,Dong:Science327:1501,Wang:Science342:1499,Yang:Science347:60}.
In these systems, the resonance is associated with the formation of
transiently stable transition-state structures. High-resolution anion photodetachment spectroscopy in combination with accurate calculations has recently allowed the observation and characterization of previously unresolved reactive scattering resonances in this system \cite{Kim31072015}. Using a merged beam
approach, resonances have recently also been observed in ICSs for
Penning ionization processes at collision temperatures in the mK regime
\cite{Henson:Science338:234,Lavert-Ofir:NatChem6:332}. Even more
recently, resonances were observed in inelastic scattering processes at
energies near the thermodynamic threshold for rotational excitation of
the molecule. Using cryogenically cooled molecular species such as CO
and O$_2$, partial wave resonances were observed in the state-to-state
ICSs for inelastic collisions with He and H$_2$ target beams at energies
down to 4 K
\cite{Chefdeville:PRL109:023201,Chefdeville:Science341:06092013,Bergeat:NatChem7:349}.

The measurement of DCSs at scattering resonances remains a largely
unexplored frontier, in particular for species other than H and D atoms.
At low collision energies, the recoil velocities of the scattered
molecules are extremely small, and it has proven a daunting task to
obtain the angular resolution required to resolve any deflection
structure. Here, we report the measurement of DCSs at partial wave
resonances for inelastic collisions between fully state-selected NO
radicals [$X\,^2\Pi_{1/2}, v=0, j=1/2, f$ \cite{Note:1}, referred to
hereafter as (1/2$f$)] and He atoms in a crossed beam experiment. We
combined Stark deceleration and velocity map imaging to probe scattering
resonances in the state-to-state and parity-resolved DCSs at energies
between 13 and 19 cm$^{-1}$, with a spectroscopic energy resolution of
0.3 cm$^{-1}$. The high resolution afforded by the Stark decelerator
allowed us to observe structure in the very small velocity mapped
scattering images, directly reflecting the DCSs. Distinct variations in
the DCSs were observed as the collision energy was tuned over the
resonances. At off-resonance energies, the DCSs were dominated by
quantum diffraction oscillations, whereas additional strong forward and
backward scattering was found
at the resonance energies. We developed a theoretical approach similar
to Feshbach-Fano partitioning \cite{fano:61,brems:02} to disentangle the
resonance and background contributions to the DCSs, and we directly
revealed the incoming and outgoing waves that characterize the resonances and
the background.

Collisions between NO and rare gas atoms represent a seminal class of
systems in rotational energy transfer, since scattering of open-shell radical species
such as NO and OH plays an important role in gas-phase chemical
kinetics, combustion, and astrochemistry. Collisions involving these
radicals are governed by multiple potential energy surfaces (PESs), parity selection and propensity
rules that are foreign to molecules such as CO and
O$_2$\cite{Kirste:Sience338:1060}. We found excellent agreement with the
DCSs predicted by \emph{ab initio} quantum mechanical close-coupling (QM CC) scattering
calculations based on accurate NO-He PESs.

We used a crossed molecular beam apparatus with a 45$^{\circ}$
crossing angle \cite{Note:SI}. A packet of velocity-controlled NO (1/2$f$) radicals,
with a velocity spread of 2.1 m/s \cite{Note:3} and an angular spread of
1.5 mrad, was produced using a 2.6-m long Stark decelerator. The beam of
He atoms with a mean velocity between 400~m/s and 480~m/s, a velocity spread of 4.3~m/s
and an angular spread of 4.8~mrad was produced by cooling an Even-Lavie
valve to cryogenic temperatures. By tuning the NO velocity between
350~m/s and 460~m/s with the Stark decelerator, the collision energy was
varied between 13 cm$^{-1}$ and 19 cm$^{-1}$ with an energy
resolution of 0.3 cm$^{-1}$. The scattered NO radicals were
state-selectively detected using a two-color laser ionization scheme and
velocity mapped on a two-dimensional detector.

We studied inelastic collisions that excite the NO (1/2$f$) radicals into either
the (3/2$e$) or the (5/2$f$) state. The (5/2$f$) channel opens within
the experimentally accessible energy range at 13.4 cm$^{-1}$, and we
measured the threshold behavior in the ICS for this channel to calibrate
both the collision energy and energy resolution (shown in Fig. 1A). We
observed a plateau just above threshold and a clear peak around 18
cm$^{-1}$, which were well reproduced by the theoretical ICS convoluted
with the experimental resolution of 0.3 cm$^{-1}$ \cite{Note:SI}. These
features were attributed to a narrow (labeled II) and a broader
resonance (labeled III) in the theoretical ICS. The theoretical ICS for
the (3/2$e$) state (shown in Fig. 1B for comparison) showed a clear
resonance (labeled I) at a slightly lower energy than resonance II.

For both inelastic channels, scattering images were measured at selected
energies (Fig. 2). Depending on the energy and the final state probed,
the diameter of the low collision energy images ranged from only a few
m/s to about 60~m/s. Note that the diameter of the (5/2$f$) image
approached zero as the energy approached the thermodynamic threshold of
the channel, effectively imaging the center of mass velocity of the
NO-He pair. Despite their small sizes, distinct structure in the images
could clearly be discerned. At higher energies, as illustrated by the
additional images probed at 45.0~cm$^{-1}$, the DCSs of both channels
feature a rich diffraction pattern that extends from forward to backward
scattering. Each diffraction peak in the DCS transformed into a stripe
in the image due to the detection method employed in the experiment. At low collision energies the number of diffraction
peaks was reduced, and an additional pattern arose in the vicinity of
the resonances. As the energy was varied in small energy steps over the
resonances, a strong variation in the angular distribution featuring
pronounced forward and backward scattering was observed.

To compare our findings with theoretical predictions, we simulated for
each energy the image expected from the kinematics of the
experiment and the DCS predicted by high-level QM CC calculations \cite{Note:SI}. Both the
experimental and simulated images were then analyzed to yield the
angular scattering distribution, reflecting the DCS convoluted with the
experimental energy and angular resolution \cite{Note:SI}. In general,
excellent agreement was found between the experimental and simulated
scattering distributions, although at some energies the relative
intensities of the observed features differed from the simulated
intensities \cite{Note:6}.

To interpret our results, we first analyzed the partial wave composition
of the scattering cross sections, as well as the scattering wavefunctions \cite{Note:SI}. We found that for the
energies probed, the entrance channel is governed by partial waves up to
$\ell_{\rm in}=$ 8. At the resonance energies, however, a single
resonant partial wave $\ell_{\rm res}$ becomes involved in a quasi-bound
state, which causes the ICS to rise significantly above the background.
We found that resonances I and II are associated with a quasi-bound state
of the He-NO(5/2$f$) complex at 14.7 cm$^{-1}$ dominated by $\ell_{\rm
res} = 5$, whereas resonance III originates from a He-NO(5/2$f$)
quasi-bound state at 17.9 cm$^{-1}$ governed by $\ell_{\rm res} = 6$
\cite{Note:SI}.

For an inelastic process, the partial wave quantum number $\ell$ is not
conserved throughout the collision. The anisotropic interaction
potential couples rotational states of the NO molecule with different
quantum numbers $j$ and waves with different values of $\ell$, and
determines how the incoming waves $\ell_{\rm in}$ are transformed into
the outgoing waves $\ell_{\rm out}$ during the collision. At the
resonance energies, the quasi-bound He-NO(5/2$f$) complexes with the
well-defined values of $\ell_{\rm res}=5$ (resonances I and II) or 6
(resonance III) decomposed to form an NO radical in either the (3/2$e$)
state (resonance I) or the (5/2$f$) state (resonances II and III). When
the resonant (5/2$f$) state decays into the lower (3/2$e$) state, as for
resonance I, not only $j$ but also $\ell$ changes and $\ell_{\rm out}$
ranges from 3 to 7 \cite{Note:SI}. When the final NO state is the same
as the resonant state, as for resonances II and III, the dominant value
of $\ell_{\rm out}=5$ or 6 is the same as $\ell_{\rm res}$. At energies
just above the threshold of the (5/2$f$) inelastic channel, where almost
all kinetic energy has been transferred into rotational energy of the NO
molecule, the (5/2$f$) state can only decay with $\ell_{\rm out} = 0$
($s$-wave), $\ell_{\rm out} = 1$ ($p$-wave) and $\ell_{\rm out} = 2$
($d$-wave). The observed DCS at 13.8 cm$^{-1}$ was dominated by $p$-wave
scattering with small contributions from $s$- and $d$-waves
\cite{Note:SI}.

At the resonance energies, the DCSs contain unique information on the
partial wave dynamics of the collision process, i.e., the relation
between $\ell_{\rm in}$, $\ell_{\rm res}$, and $\ell_{\rm out}$. If the
scattering were purely determined by a resonance without any background,
the scattering matrix $\mathbf{S}$ would be given by the 
Breit-Wigner formula, originally developed for neutron scattering in
1936 and nowadays frequently used to describe scattering processes in
high energy particle physics
\cite{Erlewein:Zphysik211:35,Sheen:IJQC9:817}. However, in most cases
and also in our experiments, the observed ICSs and DCSs result from an
interference between resonance and background contributions. We
disentangled these contributions for each of the resonances I, II, and
III by applying a theoretical analysis similar to Feshbach-Fano
partitioning \cite{fano:61,brems:02}. We wrote the energy-dependent
multichannel scattering matrix as \cite{taylor:72}
\begin{equation}
\label{eq:ff}
\mathbf{S}(E) = \mathbf{S}^{\rm bg}(E) \, \mathbf{U}^{\rm res}(E)
\end{equation}
where the background contribution $\mathbf{S}^{\rm bg}(E)$ is a slowly
varying function of the collision energy $E$ and the
resonance contribution is given by the Breit-Wigner formula
\begin{equation}
\label{eq:bw}
\mathbf{U}^{\rm res}(E) = \mathbf{I}-\frac{2 i \mathbf{A}}{E-E_{\rm res} + i\Gamma}
\end{equation}
$\mathbf{I}$ is the unit matrix, $E_{\rm res}$ is the energy of
the resonance, $\Gamma$ its width (or inverse lifetime), and the
complex-valued matrix elements $A_{\alpha\beta}=a_\alpha a^*_\beta$
contain the partial widths $a_{\alpha}$ obeying the relation
$\sum_\alpha \left|a_\alpha\right|^2 = \Gamma$. The idea associated with
the Breit-Wigner formula is that in the complex energy plane, where the
bound states correspond to poles of the $S$-matrix on the negative real
energy axis, resonances are represented by poles below the positive real
axis at positions $E_{\rm res}-i \Gamma$. By analyzing the energy
dependence of the $S$-matrix elements in the range of each resonance with
an algorithm described in the Supplementary Material \cite{Note:SI},
we could determine the parameters $E_{\rm res}$, $\Gamma$, and
$a_{\alpha}$. Then, we separated the resonance contributions to the
scattering matrix $\mathbf{S}(E)$ from the background and applied
the usual expressions \cite{lester:76} to compute the ICS and DCS from
the $S$-matrix, with or without resonance contributions.
Figure~3 shows the results for the final (5/2$f$) state of NO, where the
effects are most pronounced. The upper panel in this figure demonstrates
that the peaks in the ICS corresponding to both resonances
II and III vanish when we only included the background contribution. The
lower part of this figure illustrates the effect of the resonances on
the DCS at energies close to these resonances. The background
contributions (dashed lines) show the usual pattern of diffraction
oscillations,
which are most pronounced for small scattering angles and decrease in
amplitude for larger angles. The effect of the resonance contributions
is substantial; they lead to additional strong scattering near the
forward and backward directions. Figure~3 also shows a comparison of
measured and simulated images at 14.8, 17.1, and 18.2 cm$^{-1}$. The
simulated images were based on DCSs calculated in this energy
range, with or without resonance contributions, by taking into account
the experimental energy resolution of 0.3 cm$^{-1}$.
Clearly, the experimental images show much better agreement with
the simulations when the full DCS is taken into account.

Our theoretical analysis demonstrated that the resonances strongly
affect the nature of the DCSs, and allowed us to disentangle normal
diffraction oscillations from the resonance fingerprints in the DCSs.
The DCSs measured for collision energies in the range of the resonances
agreed very well with the DCSs obtained from the \textit{ab initio}
calculations, but only when the contributions from the resonances were
fully included. This directly confirmed that our experiment indeed
images the resonance fingerprints in the DCS.

Our joint experimental and theoretical study showed that scattering
resonances in state-to-state cross sections can now be probed with
spectroscopic resolution, even for benchmark and chemically relevant
systems that involve open-shell species such as NO. DCSs measured at the
resonance energies, combined with a theoretical analysis, provided
detailed information on the multichannel scattering process and
explicitly revealed the effects of the resonances. The theoretical
method developed to separate the resonant contributions to the ICSs and
DCSs from the background will also be applicable to other systems where
scattering resonances occur. 

\nocite{Onvlee:PCCP16:15768,Yan:RSI84:023102,Even:AiC2014:636042,Townsend:RSI74:2530,Naulin:JPCA113:14447,Zastrow:NatChem6:216,Scharfenberg:EPJD65:189,Klos:JCP112:2195,Gijsbertsen:JCP125:113112}
\bibliography{string,scibib}

\begin{thebibliography}{10}
\bibitem{Note:4}
The partial wave quantum number $\ell$ is the quantum mechanical analogue of
  the classical angular momentum $L = \mu v b$, where $\mu$ is the reduced
  mass, $v$ is the relative velocity of the colliding particles, and $b$ is the
  impact parameter defined as the distance of closest approach in the absence
  of any interaction.
\bibitem{Note:5}
Single partial wave collisions are only found at temperatures approaching zero
  kelvin, where the scattering is governed by the wave with the lowest possible
  value for $l$. Collisions between ultracold atoms and molecules near quantum
  degeneracy, for instance, are exclusively governed by partial waves with
  $l=0$ (s-wave) or $l=1$ (p-wave).
\bibitem{Note:1}
The labels $X^2\Pi_{1/2}, v$, and $j$ indicate the electronic state, the
  vibrational state and rotational state of the NO radical, respectively. Each
  rotational state of NO possesses two near-degenerate $\Lambda$-doublet
  components, with symmetry labels $e$ (lower component) and $f$ (upper
  component), that refer to the total parity of the electronic wavefunction,
  exclusive of rotation.
\bibitem{Note:6}
The calibration of the mean experimental collision energy has an uncertainty of
  $\pm$~0.15~cm$^{-1}$. Small differences between experimental and simulated
  images are found in particular at energies where the DCS varies rapidly as a
  function of collision energy. In addition, the experiment seemed to
  systematically underestimate the scattering intensity at forward scattering.
\end{thebibliography}

\bibliographystyle{Science}

{\bf Acknowledgement} This work is part of the research programme of the
Foundation for Fundamental Research on Matter (FOM), which is supported
financially by the Netherlands Organization for Scientific Research
(NWO). S.Y.T.v.d.M. acknowledges further support from NWO via a VIDI and
a TOP grant, and from the European Research Council via a Starting
Grant. We thank Andr\'e van Roij, Chris Berkhout, Niek Janssen, and
Peter Claus for expert technical support. We thank Jacek K{\l}os for
providing us with his NO-He PESs. The authors declare no competing
financial interests. Supplementary Material accompanies this paper.\\

{\bf Supplementary Materials}\\*
Materials and Methods\\*
Figs. S1 to S10\\*
Table S1\\*
References (28-36)\\*
Movies S1-S2\\*

\begin{figure}
 \centering
 \includegraphics[width=0.6\columnwidth]{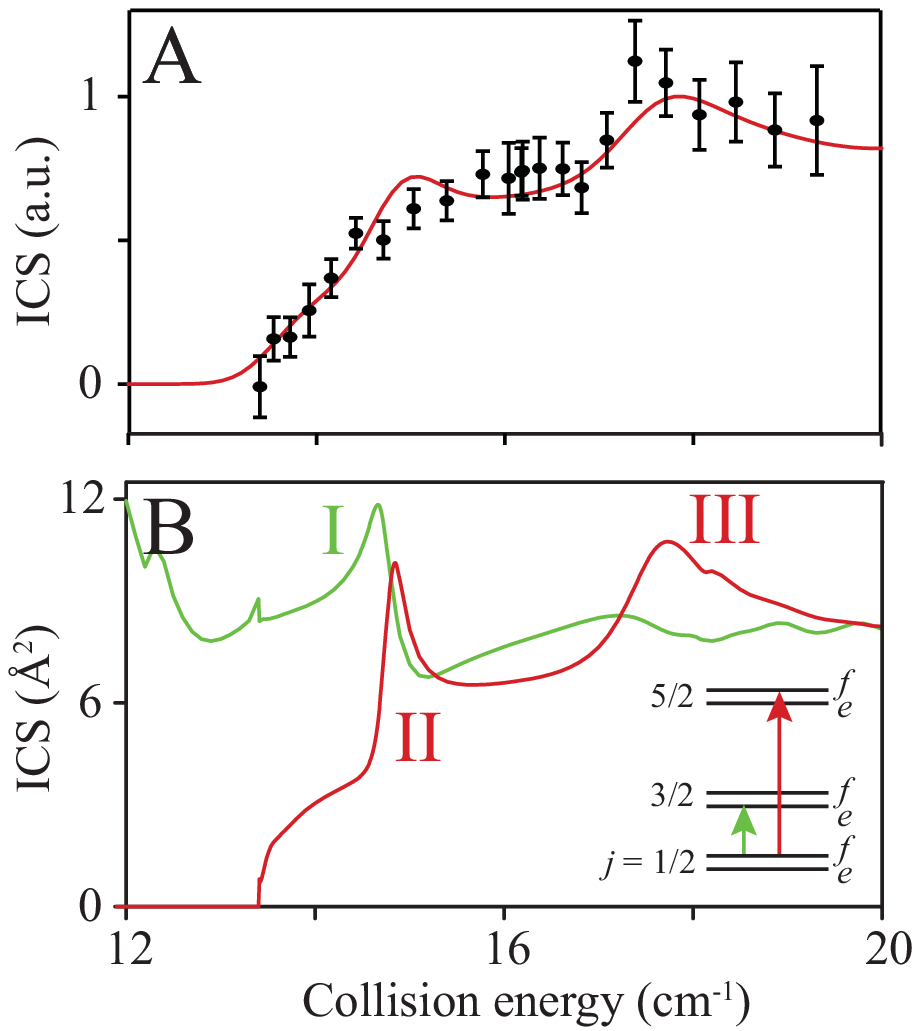}
\caption{}
\end{figure}

\begin{figure}
 \centering
 \includegraphics[width=0.9\columnwidth]{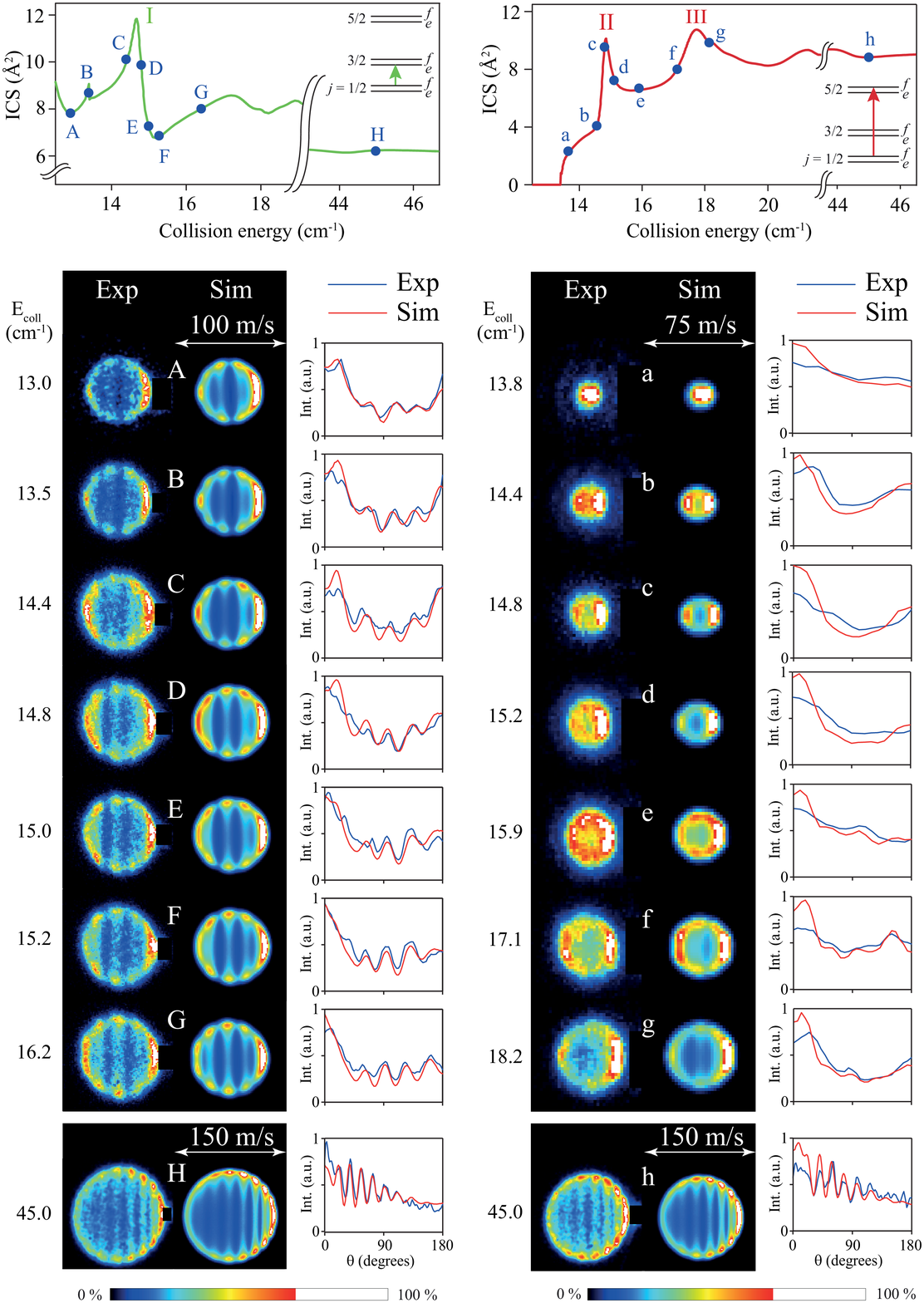}
\caption{}
\end{figure}

\begin{figure}
 \centering
 \includegraphics[width=0.9\columnwidth]{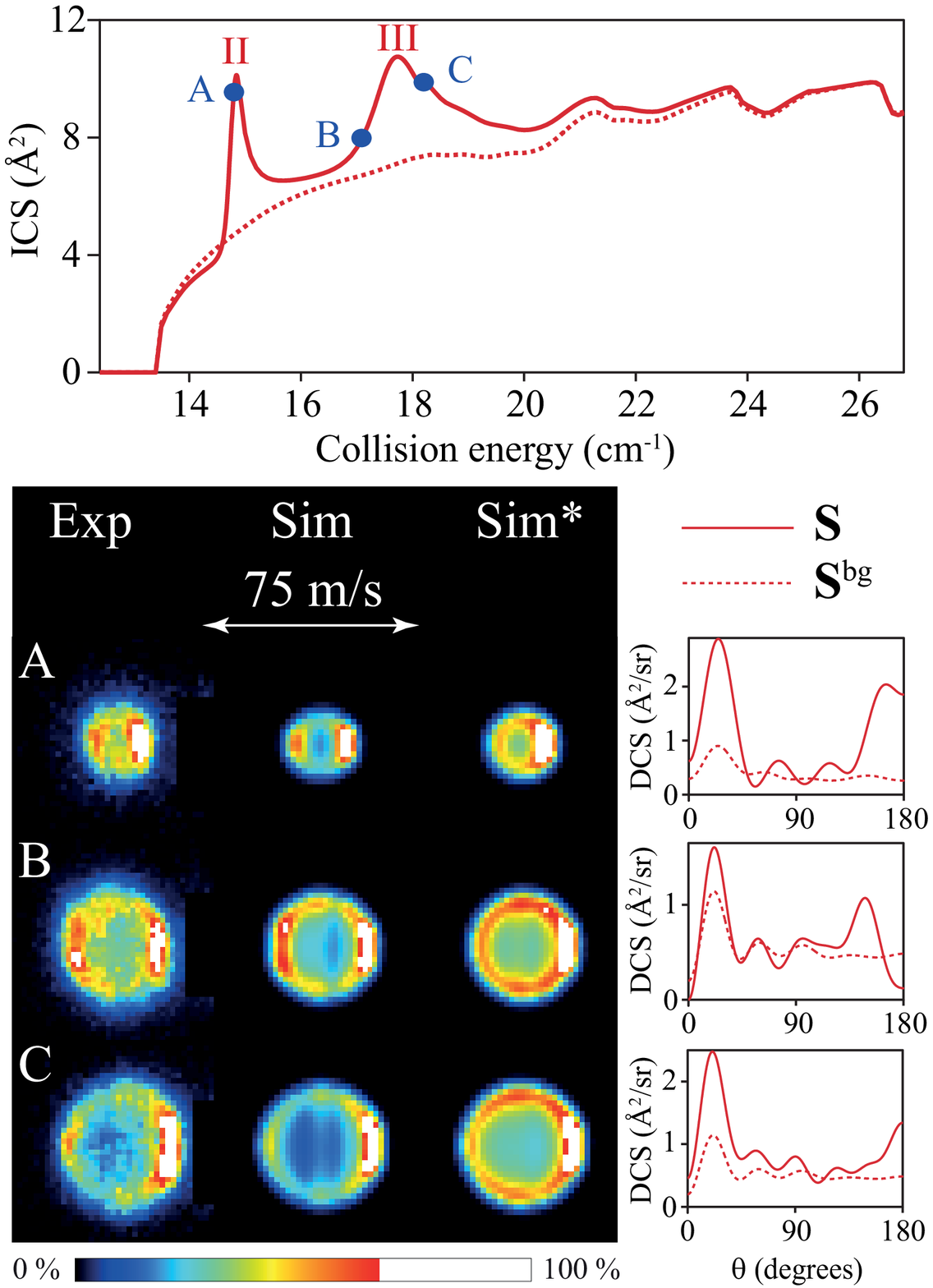}
\caption{}
\end{figure}

\newpage
\noindent
Figure Captions:\\[2ex]

\noindent
Figure 1: Collision energy dependence of the integral cross section for
rotational excitation of NO radicals by He atoms. (A) Comparison between
measured (data points with error bars) and calculated (solid curve)
state-to-state inelastic scattering cross sections for excitation into
the (5/2$f$) state. Experimental data in a.u., arbitrary units. Each data point is averaged over 1000 laser shots with the He and NO beams overlapping (collision signal), and 1000 laser shots with the NO beam only (reference signal). Vertical
error bars represent statistical uncertainties at 95\% of the confidence
interval. The calculated cross section was convoluted
with the experimental energy resolution of 0.3~cm$^{-1}$. (B) Calculated
state-to-state integral cross sections for excitation into the (3/2$e$) state (green curve)
and (5/2$f$) state (red curve). (Inset) Schematic energy level diagram and
inelastic excitation scheme of NO. \\[2ex]

\noindent
Figure 2: Experimental (Exp) and simulated (Sim) ion images at selected
collision energies as indicated in the top panels. Left panels: (1/2$f$)
$\rightarrow$ (3/2$e$) inelastic collisions. Right panels: (1/2$f$)
$\rightarrow$ (5/2$f$) inelastic collisions. The images are presented
such that the relative velocity vector is oriented horizontally, with
the forward direction on the right side of the image. Small segments of
the images around forward scattering are masked due to imperfect state
selection of the NO packet. The angular scattering distributions as
derived from the experimental (blue curves) and simulated (red curves)
images are shown for each channel and collision energy. \\[2ex]

\noindent
Figure 3: Effect of resonances II and III on the cross sections for inelastic $(1/2f)
\rightarrow (5/2f)$ NO-He scattering. Integral cross sections are shown above, differential below. Solid lines represent the complete
theoretical ICSs and DCSs, dashed lines the cross sections obtained when
only the scattering matrix $\mathbf{S}^{\rm bg}$ in Eq.~(\ref{eq:ff}) is
included for resonances I, II, and III. The lower panels show the measured
(Exp) and simulated images based on either the complete DCSs (Sim) or
the DCSs computed with the scattering matrix $\mathbf{S}^{\rm bg}$ only
(Sim*) for collision energies of (A) 14.8 cm$^{-1}$, (B) 17.1 cm$^{-1}$,
and (C) 18.2 cm$^{-1}$. Fig. S10 shows resonance I in the $(1/2f)
\rightarrow (3/2e)$ cross sections.

\end{document}